\documentclass[10pt,english,showpacs, secnumarabic, showkeys]{revtex4}
\usepackage[T1]{fontenc}
\usepackage[latin9]{inputenc}
\usepackage{amsmath}
\usepackage{esint}
\usepackage{babel}

\begin{document}

\title{The standard model and the four dimensional superstring}

\author{B. B. Deo}

\affiliation{Department of Physics, Utkal University, Bhubaneswar-751004, India}

\email{bdeo@iopb.res.in}

\author{P. K. Jena}

\affiliation{72, Dharma Vihar, Khandagiri, Bhubaneswar-751030, India}

\email{prasantajena@yahoo.com}

\date{31 Jan, 2011}

\begin{abstract}
Starting from the Nambu-Goto bosonic string, a four dimensional superstring
model is constructed using the equivalence of one boson to two Majorana-Weyl
fermions. The conditions of anomaly cancellation in a `heterotic'
string theory lead to the correct result and is found to be consistent
with the requirements of the standard model.
\end{abstract}



\maketitle
String theories are promising candidates for providing an unified
theory of the physical world as they offer the possibility of explaining
the spectra of the chiral matter field interacting through the gauge
and gravitational forces. One can work with the Nambu-Goto\citet{key-1,key-2}
action involving the string coordinates $X^{\mu}\left(\sigma,\tau\right)$
in the world sheet $\left(\sigma,\tau\right)$ whose dimension turns
out to be 26. The most pressing problem faced by the string theory
is that of breaking a 26 dimensional string theory down to four dimensions.
Unless such a dimensional reduction is achieved, the theory cannot
have any connection with the observable quantities of the real physical
world.

The Nambu-Goto bosonic string in the world sheet $\left(\sigma,\tau\right)$
in 26 dimensions is described by the action\citet{key-1}

\begin{equation}
S_{B}=-\frac{1}{2\pi}\int\, d^{2}\sigma\,\partial_{\alpha}X^{\mu}\left(\sigma,\tau\right)\partial^{\alpha}X_{\mu}\left(\sigma,\tau\right),\,\,\,\,\,\mu=0,1,\cdots,25.\label{eq:1}\end{equation}
This action can be rewritten as the sum of (i) the action for four
bosonic coordinates $X^{\mu},\,\,\mu=0,1,2,3$ and (ii) the action
for 44 fermions having $SO\left(44\right)$ symmetry. This decomposition
is possible by using Mandelstam's proof of the equivalence between
one boson to two fermions, in the infinite volume limit, in $\left(1+1\right)$
dimensional field theory. These fermions are taken to be the Majorana-Weyl
fermions. The corresponding fermionic-bosonic action takes the form\citet{key-3,key-4,key-5}

\begin{equation}
S_{FB}=-\frac{1}{2\pi}\int\left[\partial_{\alpha}X^{\mu}\partial^{\alpha}X_{\mu}-i\sum_{j=1}^{6}\bar{\psi}^{\mu,\, j}\rho^{\alpha}\partial_{\alpha}\psi_{\mu,\, j}+i\sum_{k=1}^{5}\bar{\phi}^{\,\mu,\, k}\rho^{\alpha}\partial_{\alpha}\phi_{\mu,\, k}\right]\, d^{2}\sigma,\label{eq:2}\end{equation}
where $\rho^{0}=\left(\begin{array}{cc}
0 & -i\\
i & 0\end{array}\right),\,\,\rho^{1}=\left(\begin{array}{cc}
0 & i\\
i & 0\end{array}\right),\,\,\bar{\psi}=\psi^{\dagger}\rho^{0}$ and $\bar{\phi}=\phi^{\dagger}\rho^{0}$. The Dirac operators $\rho^{\alpha}\partial_{\alpha}$
are hermitian since the matrices $\rho^{0}$ and $\rho^{1}$ are imaginary.
Further, the fermionic fields are $\psi^{\mu,j}=\psi^{(+)\,\mu,j}+\psi^{(-)\,\mu,j}$
and $\phi^{\mu,k}=\phi^{(+)\,\mu,k}+\phi^{(-)\,\mu,k}$. Out of the
44 fermions, there are $6\times4=24$ ``neutrinos'' of one type and
$5\times4=20$ ``neutrinos'' of another type which differ in their
``neutrino-like'' phase. This phase difference leads to the opposite
signs in the 2nd and 3rd terms in the action of equation (\ref{eq:1}).
This action is invariant under $SO(6)\times SO(5)$ as well as under
$SO(3,1)$. It is also invariant under the supersymmetric transformation

\begin{equation}
\delta X^{\mu}=\bar{\epsilon}\left(e^{j}\psi_{j}^{\mu}-e^{k}\phi_{k}^{\mu}\right),\,\,\delta\psi^{\mu,j}=-i\epsilon e^{j}\rho^{\alpha}\partial_{\alpha}X^{\mu},\,\,\delta\phi^{\mu,k}=i\epsilon e^{k}\rho^{\alpha}\partial_{\alpha}X^{\mu},\label{eq:3}\end{equation}
where $\epsilon$ is a constant anticommuting spinor. The $e^{j}$
are arrays of 11 numbers with one `1' in the $j$ th position and
the rest equal to zero. Similarly, the $e^{k}$ are arrays of 11 numbers
with only one `-1' in the $k$ th position and the rest equal to zero.
The $e^{j}$ and $e^{k}$ satisfy $e^{j}e_{j}=6$ and $e^{k}e_{k}=5.$
The commutator of two supersymmetric transformations gives a world
sheet transformation. Further, the fermionic combination $\Psi^{\mu}=\left(e^{j}\psi_{j}^{\mu}-e^{k}\phi_{k}^{\mu}\right)$
is the superpartner of $X^{\mu}$.

In order to cancel the anomalies and isolate the natural ghosts, the
Faddeev-Popov ghosts $b$ and $c$ are introduced\citet{key-6,key-7}.
The corresponding $FP$ ghost action is\citet{key-5},

\begin{equation}
S_{FP}=\frac{1}{\pi}\int d^{2}\sigma\,\left(c^{+}\partial_{-}b_{++}+c^{-}\partial_{+}b_{--}\right),\label{eq:4}\end{equation}
with the anticommutation relations 

\begin{equation}
\left\{ b_{\pm\pm}\left(\sigma,\tau\right),c^{\pm}\left(\sigma^{\prime},\tau\right)\right\} =2\pi\delta\left(\sigma-\sigma^{\prime}\right).\label{eq:5}\end{equation}
The mode expansions for the Faddeev-Popov ghosts are given by

\begin{equation}
c^{\pm}=\sum_{n=-\infty}^{\infty}c_{n}e^{-in\left(\tau\pm\sigma\right)},\,\,\,\,\,\, b_{\pm\pm}=\sum_{n=-\infty}^{\infty}b_{n}e^{-in\left(\tau\pm\sigma\right)},\label{eq:6}\end{equation}
which lead to the canonical anticommutation relations

\begin{equation}
\left\{ c_{m},b_{n}\right\} =\delta_{m,-n},\,\,\,\,\left\{ c_{m},c_{n}\right\} =\left\{ b_{m},b_{n}\right\} =0.\label{eq:7}\end{equation}
Using the mode expansion and the energy-momentum tensor, obtained
from the action, we get the Virasoro generators for the ghosts $G$
as

\begin{equation}
L_{m}^{G}=\sum_{n=-\infty}^{\infty}\left(m-n\right)b_{m+n}c_{-n}-a\delta_{m,-n},\label{eq:8}\end{equation}
where the normal ordering constant equals 1. The action $S_{FB}$
of equation(\ref{eq:2}) describes the left moving sector.

In the right moving sector we have, in addition, 11 superconformal
ghosts which appears puzzling. But the light-cone gauge is ghost free.
The light-cone fields are

\begin{equation}
\psi_{j}^{\pm}=\frac{1}{\sqrt{2}}\left(\psi_{j}^{0}\pm\psi_{j}^{3}\right),\,\,\,\phi_{k}^{\pm}=\frac{1}{\sqrt{2}}\left(\phi_{k}^{0}\pm\phi_{k}^{3}\right),\label{eq:9}\end{equation}
with anticommutators having negative sign,

\begin{equation}
\left\{ \psi_{j}^{+}\left(\sigma\right),\psi_{j^{\prime}}^{-}\left(\sigma^{\prime}\right)\right\} =-\delta_{jj^{\prime}}\delta\left(\sigma-\sigma^{\prime}\right),\,\,\left\{ \phi_{j}^{+}\left(\sigma\right),\phi_{j^{\prime}}^{-}\left(\sigma^{\prime}\right)\right\} =-\delta_{jj^{\prime}}\delta\left(\sigma-\sigma^{\prime}\right).\label{eq:10}\end{equation}
The total energy-momentum tensor, which comes from the $\left(0,3\right)$
coordinates is given by

\begin{equation}
T^{gh}(z)=\frac{i}{2}\left(\psi^{0j}\partial_{z}\psi_{0j}+\psi^{3j}\partial_{z}\psi_{3j}\right)+\frac{i}{2}\left(\phi^{0j}\partial_{z}\phi_{0j}+\phi^{3j}\partial_{z}\phi_{3j}\right).\label{eq:11}\end{equation}
The corresponding correlation function is

\begin{equation}
\langle T^{gh}(z),T^{gh}(\omega)\rangle=\frac{11}{2}\frac{1}{\left(z-\omega\right)^{4}}+\cdots\label{eq:12}\end{equation}
These ghosts contribute 11 to the central charge whereas the field
components are only $\frac{D}{4}=1.$ The appropriate action is taken
to be

\begin{equation}
S=-\frac{1}{2\pi}\int d^{2}\sigma\,\left[\sum_{\mu=0}^{3}\partial_{\alpha}X^{\mu}\partial^{\alpha}X_{\mu}-i\sum_{\mu=1,2}\left(\bar{\psi}^{\mu,\, a}\rho^{\alpha}\partial_{\alpha}\psi_{\mu,\, a}-\bar{\phi}^{\,\mu,\, b}\rho^{\alpha}\partial_{\alpha}\phi_{\mu,\, b}\right)+i\sum_{\mu=0,3}\left(\bar{\psi}^{\mu,\, a}\rho^{\alpha}\partial_{\alpha}\psi_{\mu,\, a}-\bar{\phi}^{\,\mu,\, b}\rho^{\alpha}\partial_{\alpha}\phi_{\mu,\, b}\right)\right].\label{eq:13}\end{equation}
The anomaly cancellation is automatic in this case.

The heterotic string\citet{key-8} has the advantage of having two
independently moving sectors of closed strings, namely the right moving
sector propagating as function of $\left(\sigma+\tau\right)$ and
the the left moving sector as a function of $\left(\sigma-\tau\right)$
. The term ``heterosis'' , meaning `hybrid vigor' makes fundamental
use of this splitting. It has been shown\citet{key-8,key-9,key-10}
that the heterotic string has no tachyons, no anomalies and is finite
to one loop as considered here.

The number of compactified dimensions in the left moving sector is
$\left(26-D\right)=22$. In this sector, the three contributions to
anomaly are as follows\citet{key-9}.

\begin{equation}
\text{Left\, movers}:\,\,\, X_{\mu}\longrightarrow D;\,\, b,c\,\,\,\text{ghosts}\longrightarrow26;\,\,\,\,\psi^{a}\longrightarrow\frac{N}{2}.\label{eq:14}\end{equation}
The sum of the three contributions to anomaly must vanish so that
the number of fermions in the set $\psi^{a}$, for $D=4,$ is equal
to 

\begin{equation}
N_{L}=2(26-D)=44.\label{eq:15}\end{equation}
On the other hand, the contributions to the anomaly from the right
moving sector are

\begin{equation}
\text{Right\, movers}:\,\, X_{\mu},\,\psi_{\mu}\longrightarrow\frac{5}{4}D;\,\, b,c,\beta,\gamma\,\,\longrightarrow-26+11=-15;\,\,\,\psi^{a}\longrightarrow\frac{N}{2}.\label{eq:16}\end{equation}
The condition for cancellation of anomaly in this sector, for $D=4$,
leads to the number of fermions as

\begin{equation}
N_{R}=3\left(10-\frac{5}{6}D\right)=20.\label{eq:17}\end{equation}

We now construct the super Kac-Moody and Virasoro algebras for compactified
models. Our aim is to find the superconformal anomaly of the super
Virasoro ghosts for the various fermionic representations\citet{key-11,key-12}.
Using the fermionic fields $\psi^{a}$, it is possible to construct
representations of the super Kac-Moody generators $J_{B}^{a},J_{F}^{a}$
and the super Virasoro generators $T_{B},T_{F}$. With the Grassman
variable $\theta,$ the super Kac-Moody generators are 

\begin{equation}
J^{a}(z,\theta)=J_{F}^{a}+\theta J_{B}^{a},\label{eq:18}\end{equation}
where 

\begin{equation}
J_{F}^{a}=\sqrt{k}\psi^{a},\,\,\,\, J_{B}^{a}=-\frac{i}{2}f^{abc}\psi^{b}\psi^{c}.\label{eq:19}\end{equation}
Here, $f^{abc}$ are the structure constants and $k$ is the level
number. The anomaly term is found to be 

\begin{equation}
k=c_{2}(G),\label{eq:20}\end{equation}
where $c_{2}(G)$ is the value of the quadratic conserved operator
for the group $G$, which for the standard model is $SU(n)$ so that
$c_{2}(G)=n$. We also have the relations 

\begin{equation}
T(z,\theta)=T_{F}+\theta T_{B},\label{eq:21}\end{equation}
with

\begin{equation}
T_{B}=-\frac{1}{2}\psi^{a}\partial\psi_{a},\,\,\,\,\,\,\text{and}\,\,\,\,\,\, T_{F}=-\frac{i}{12\sqrt{k}}f^{abs}\psi^{a}\psi^{b}\psi^{c}.\label{eq:22}\end{equation}
The conformal anomaly for this representation of super Virasoro algebra
is

\begin{equation}
c=\frac{1}{2}d(G),\label{eq:23}\end{equation}
where $d(G)$ is the dimension of the group $G$. So, for $SU(n)$,
we have

\begin{equation}
c=\frac{1}{2}(n^{2}-1).\label{eq:24}\end{equation}
If we have a super Kac-Moody generator $\tilde{J}$, such that 

\begin{equation}
J_{B}^{a}=\tilde{J}^{a}-\frac{i}{2}f^{abc}\psi^{b}\psi^{c}\,\,\,\,\,\,\text{and\,}\,\,\,\,\, J_{F}^{a}=\sqrt{k}\psi^{a},\label{eq:25}\end{equation}
the expressions for $T_{B}$ and $T_{F}$ become

\begin{equation}
T_{B}=-\frac{1}{2}\psi^{a}\partial\psi_{a}+\frac{1}{2k}\tilde{J}^{a}\tilde{J}^{a},\label{eq:26}\end{equation}

\begin{equation}
T_{F}=-\frac{i}{12\sqrt{k}}f^{abs}\psi^{a}\psi^{b}\psi^{c}+\frac{1}{2\sqrt{k}}\psi^{a}\tilde{J}^{a}.\label{eq:27}\end{equation}
If $\tilde{k}$ is added to the level number $k$ , so that $k=\tilde{k}+c_{2}(G),$
the conformal anomaly becomes

\begin{equation}
c=\frac{1}{2}d(G)+d(G)\frac{k-c_{2}(G)}{k}.\label{eq:28}\end{equation}
The first term comes from the $\psi^{a}$ field and the second term
from the energy-momentum tensor. Since $d(G)=n^{2}-1$ and $c_{2}(G)=n$,
we find that

\begin{equation}
\hat{c}=(n^{2}-1)\left[1-\frac{2n}{3k}\right]=c\frac{2}{3}.\label{eq:29}\end{equation}

From this equation (\ref{eq:29}), we get the following.

\begin{enumerate}
\item If $G=SU(3)$, quarks are triplets, $\hat{k}\geq1$ and $\hat{c}\geq4.$
\item If $G=SU(2)$ , quarks are doublets, $\hat{k}\geq1$ and $\hat{c}\geq\frac{5}{4}.$
\item If $G=U(1)$, $\hat{c}\geq1.$
\end{enumerate}
So, the total contribution to anomaly for the Standard Model is

\begin{equation}
\hat{c}\geq\left(4+\frac{5}{3}+1\right)=6\frac{2}{3}.\label{eq:42}\end{equation}
Indeed this is exactly the result we require, i.e., $\hat{c}=6\frac{2}{3}$
or $c=10.$ 

Pursuing further, we note that our action has $SO(6)\otimes SO(5)$
symmetry. In order to descend to the Standard Model group $SU_{C}(3)\otimes SU_{L}(2)\otimes U_{Y}(1)$,
one usually introduces Higgs which break the gauge symmetry and supersymmetry.
But the use of Wilson's loops

\begin{equation}
U_{\gamma}=P\exp\left(\oint_{\gamma}A_{\mu}dx^{\mu}\right),\label{eq:30}\end{equation}
breaks the gauge symmetry while keeping the supersymmetry intact\citet{key-12,key-13}.
Here, $P$ stands for the ordering of each term with respect to the
closed path $\gamma.$ The group $SO(6)=SU(4)$ descends to $SU_{C}(3)\otimes U_{B-L}(1)$
by choosing one element of $U_{0}$ of $SU(4)$ such that $U_{0}^{2}=1$.
The element generates the permutation group $Z_{2}$, so that $\frac{SO(6)}{Z_{2}}=SU_{C}(3)\otimes U_{B-L}(1),$without
breaking supersymmetry. Similarly, $SO(5)\rightarrow SO(3)\otimes SO(2)=SU(2)\otimes U(1)$,
which gives $\frac{SO(5)}{Z_{2}}=SU(2)\otimes U(1).$Thus

\begin{equation}
\frac{SO(6)\otimes SO(5)}{Z_{2}\otimes Z_{2}}=SU_{C}(3)\otimes U_{B-L}(1)\otimes U_{R}(1)\otimes SU_{L}(2),\label{eq:33}\end{equation}
which makes an identification with the usual low energy phenomenology.
However, this is not the Standard Model as there is an additional
$U(1)$. This difficulty is overcome by reducing the rank by one as
dicussed below\citet{key-11,key-12}. We have for the three Wilson's
loops,

\begin{equation}
g\left(\theta_{1},\theta_{2},\theta_{3}\right)=\left(\frac{2\pi}{3}-2\theta_{1},\frac{2\pi}{3}+\theta_{2},\frac{2\pi}{3}+\theta_{3}\right).\label{eq:34}\end{equation}
The first Wilson loop integral vanishes as the angle integral is from
$\theta_{1}=\frac{2\pi}{9}$ to $\frac{2\pi}{3}-\frac{4\pi}{9}=\frac{2\pi}{9}$.
The second loop, for which $\theta_{2}=0$ to $2\pi-\frac{2\pi}{3}=\frac{4\pi}{3}$,
is described by a length parameter $R$. The third loop, with $\theta_{3}=0\,\,\text{to\,\,\,}2\pi-\frac{2\pi}{3}=\frac{4\pi}{3}$,
is also described by the same parameter $R$. The polar components
of the gauge fields are taken as nonzero constants so that $gA_{\theta_{2}}^{15}=\vartheta_{15}$for
$SO(6)=SU(4)$. The diagonal generator $t_{15}$ breakes the symmetry.
Similarly, for $SO(5)$, we have $g^{\prime}A_{\theta_{3}}^{10}=\vartheta_{10}^{\prime}$with
the corresponding diagonal generator being $t_{10}^{\prime}.$ The
generators of both $SO(6)$ and $SO(5)$ are $4\times4$ matrices
and the group $Z_{3}$ can be written as $T=T_{\theta_{1}}T_{\theta_{2}}T_{\theta_{3}}.$The
unbroken symmetry $SU(3)\otimes SU(2)$ is not affected since $T_{\theta_{1}}=1.$
But $T_{\theta_{2}}=\exp\left(it_{15}\int_{0}^{\frac{4\pi}{3}}\vartheta_{15}Rd\theta_{2}\right)\neq1,$breaks
the $SU(4)$ symmetry and $T_{\theta_{3}}=\exp\left(it_{10}^{\prime}\int_{0}^{\frac{4\pi}{3}}\vartheta_{10}^{\prime}R\, d\theta_{2}\right)\neq1,$breaks
the $SO(5)$ symmetry. The remaining product of $Z_{3}$ is $T_{\theta_{2}}T_{\theta_{3}}=\exp\left(i\int_{0}^{\frac{4\pi}{3}}\left(\vartheta_{10}^{\prime}t_{10}^{\prime}+\vartheta_{15}t_{15}\right)R\, d\theta_{}\right).$The
arbitrary constants $\vartheta_{15}$ and $\vartheta_{10}^{\prime}$
are chosen so as to give

\begin{equation}
\vartheta_{10}^{\prime}t_{10}^{\prime}+\vartheta_{15}t_{15}=0,\frac{3}{2R},\cdots\label{eq:40}\end{equation}
With this choice the term in the exponential in the product $T_{\theta_{2}}T_{\theta_{3}}$
becomes equal to zero or multiples of $2\pi i$. Thus, $T=U(1)$ and
the rank is reduced by one, leading to

\begin{equation}
\frac{SO(6)\otimes SO(5)}{Z_{3}}=SU_{C}(3)\otimes SU_{L}(2)\otimes U_{Y}(1).\label{eq:41}\end{equation}

Thus the absence of fermions in the 26 dimensional string theory is
not the problem. The number 26 appears over and over again in the
string theory in 4 or 10 dimensions. So, even though supersymmetry
emerges as an `accident' as we come fro 26 to 4 dimensions, the theory
gives the correct truncation of the original bosonic string theory.

\end{document}